\documentclass[final,3p,times]{elsarticle}
\usepackage{xurl}

\usepackage{amssymb,amsmath,amsfonts,mathtools}
\usepackage{amsthm}

\usepackage{graphicx}
\usepackage{graphbox}
\usepackage{algorithm}
\usepackage{algpseudocode}
\usepackage{subcaption}
\usepackage{tabularx,booktabs}
\newcolumntype{R}{>{\raggedright\arraybackslash}X}
\newcolumntype{L}{>{\raggedleft\arraybackslash}X}
\newcolumntype{Y}{>{\centering\arraybackslash}X}

\usepackage{multirow}
\usepackage{systeme}
\biboptions{sort&compress}
\usepackage{lineno}
\modulolinenumbers[5]
\usepackage[]{setspace}

\usepackage{listings}
\usepackage{color} 
\definecolor{mygreen}{RGB}{28,172,0} 
\definecolor{mylilas}{RGB}{170,55,241}

\journal{Applied Energy}

\begin{document}

\begin{frontmatter}

\title{Estimating Solar and Wind Power Production using Computer Vision Deep Learning Techniques on Weather Maps}


\author[St]{Sebastian BM Bosma\corref{cor1}}
\ead{sbosma@stanford.edu}

\author[St]{Negar Nazari}
\ead{nazari@stanford.edu}

\cortext[cor1]{Corresponding author}

\address[St]{Department of Energy Resources Engineering, Stanford University, Stanford, CA, USA}

\begin{abstract}
Accurate renewable energy production forecasting has become a priority as the share of intermittent energy sources on the grid increases. Recent work has shown that convolutional deep learning models can successfully be applied to forecast weather maps. Building on this capability, we propose a ResNet-inspired model that estimates solar and wind power production based on weather maps. By capturing both spatial and temporal correlations using convolutional neural networks with stacked input frames, the model is designed to capture the complex dynamics governing these energy sources. A dataset that focuses on the state of California is constructed, and made available as a secondary contribution of the work. We demonstrate that our novel model outperforms traditional deep learning techniques: it predicts an accurate power production profile that is smooth and includes high frequency details.
\end{abstract}

\begin{keyword}
Renewable Energy \sep
Power Forecasting \sep
Weather Maps \sep
Computer Vision \sep
Convolutional Neural Networks  \sep
Deep Learning 
\end{keyword}

\end{frontmatter}



\allowdisplaybreaks

\section{Introduction and Recent Work}
As the world attempts to mitigate severe climate change and chases its ambitious clean energy goals, renewable energy capacity continues to grow exponentially. Due to the intermittent nature of renewable power, the market supply of electricity is therefore becoming increasingly unpredictable. Simultaneously, more energy intensive industries and a larger workforce are creating sizable swings in energy demand. This fluctuating supply-demand mismatch has led to volatile electricity market prices. As a result, both energy consumers and energy providers can substantially benefit from improved short-term production, demand, and price forecasting \cite{BRANCUCCI2016valuesolarforecasting, JONSSON2010marketimpactwindforecasts}. In the bigger picture, improved forecasts can lead to lower costs of renewable energy projects \cite{hodge2018combined} and therefore higher renewable energy penetration.

As market complexity increases, the performance of traditional methods has become less reliable. Recently, data driven approaches have started to offer better alternatives \cite{ANTONANZAS2016reviewphotovoltaic,WANG2019review}. Most notably, efforts have focused on (i) predicting wind or solar plant power output using high resolution local weather data \cite{Chen19,Gensler16,Sharma2011,Zhu17}, (ii) predicting energy spot prices using market data and trends \cite{Jiang18priceLSTM,Lago18}, and (iii) predicting energy variables primarily using the variable’s own signal, sparsely distributed weather stations and other one-dimensional data streams \cite{Mashlakov2021probabiltisticForecasting, Khodayar2019GraphWindForecasting, bedi2019deep}. Although successful, these approaches have limitations. The first approach requires large high-resolution weather data sets that are globally unavailable and lead to computationally expensive solution strategies that scale poorly. The second approach fails to directly connect important market drivers, such as renewable energy supply, to market prices. The third approach resolves these shortcomings yet does not leverage the spatial information embedded in weather data. To our knowledge, few works have attempted to marry mesoscale weather information --- e.g. weather maps --- and system-scale energy production forecasting.  Most successfully, Mathe et al. \cite{mathe19pvnet} introduced a long-term recurrent convolutional network using numerical weather predictions to, in turn, predict PV production in the 24-hour and 48-hour forecast horizons. The work presented in this paper continues in that vein of thought. Other related approaches have forecasted next-day regional solar power production by using long short-term memory (LSTM) models with heat maps of current solar power production as input \cite{Chai2020robustforecastingframework, Almaghrabi21RegionalSolar}. 

Convolutional neural networks (CNNs) have been extensively applied to atmospheric science applications. Due to CNNs’ ability to find complex relationships in data series while preserving spatial information, they are inherently well suited to capture the nonlinear and spatially connected nature of weather. Popular use cases include CNN models to improve or complement numerical weather predictions \cite{Chat20,weyn2020improving,Scher18weatherclimateforecasting,rasp2020weatherbench} and CNN models that predict the occurrence of hazardous atmospheric phenomena \cite{McGovern17highimpactweather, KAMANGIR2021FogNet}. Other models have also successfully been applied to the same problem, these include DenseNets \cite{Huang_2017_CVPR}, deep belief networks \cite{Chen19}, support vector machines \cite{Jang2016}, and stacked AutoEncoders \cite{Gensler16}.

To improve forecasts of regional energy quantities such as renewable energy production, we envision a strategy that leverages accurate existing weather forecasting capabilities and computer vision deep learning techniques. Specifically, we divide the renewable energy forecasting task into two distinct problems, each with a specialized model: (i) a weather forecasting model and (ii) a power estimation model. As the prior is a well-established field, in this work we focus on the latter and develop an estimator for solar and wind power production using surface weather maps. This estimator can then be applied to forecast energy production by inputting weather forecasts maps – a global strategy similar to the one proposed in \cite{mathe19pvnet}. For the estimation task, we propose a ResNet \cite{He_2016_CVPR_ResNet} inspired model that takes a stack of five weather images as input. These five weather images correspond to the five hours leading up to the estimation time, $t$. To verify the performance of the proposed estimator, we compare its performance with respect to a linear neural network model and vary the input stack size between one and five. 

The model was developed using a data set that focused on California -- an interesting test case considering the large renewable energy market share. The renewable energy production data was obtained from the California Independent System Operator (CAISO) \cite{CaisoProductionData}, and weather maps are sourced from the National Oceanic and Atmospheric Administration (NOAA) via Google Earth Engine (GEE) \cite{WeatherData}. The collected data set is a secondary contribution of this work. 

This paper is structured as follows. We introduce the renewable power estimation problem. Next, the generation of the dataset and its properties are discussed. Subsequently, we explore the various models and training designs. The models’ results are then presented, compared, and analyzed. We conclude with several avenues for future work.

\section{The Renewable Energy Estimation Problem} \label{sec:ProblemDescription}
\begin{figure}[htbp]
\begin{center}
   \includegraphics[width=.8\linewidth]{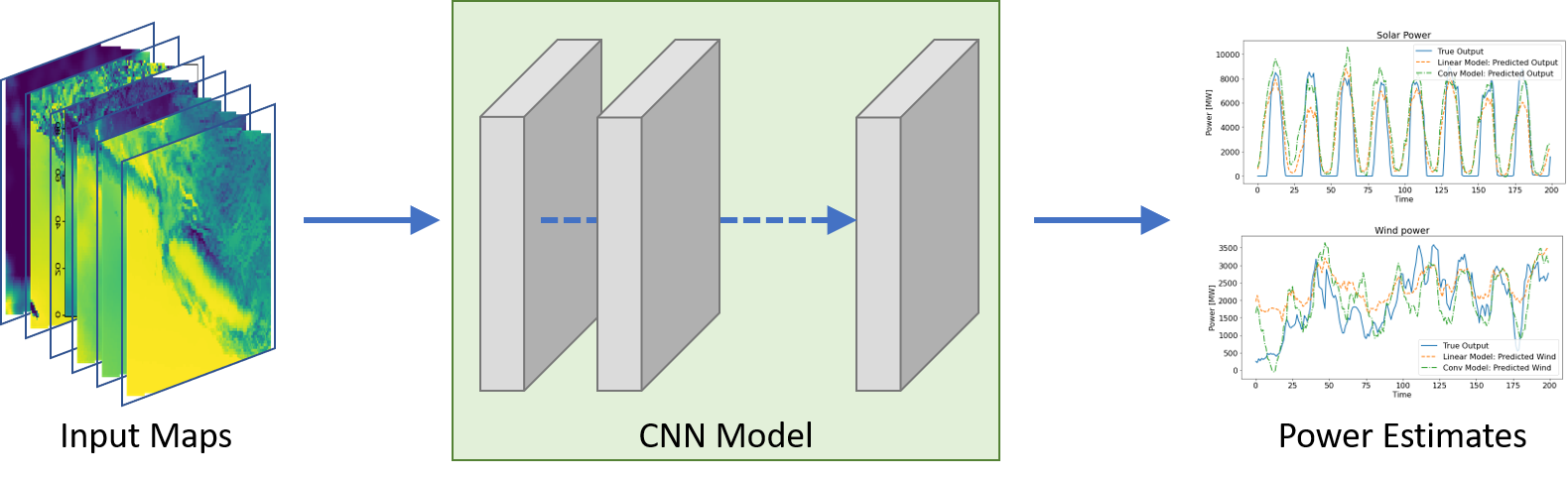}
\end{center}
   \caption{General workflow of the problem in a conventional deep learning set-up. Input maps are the six weather channels for each sample. Output power estimates are for wind and solar energy.}
\label{fig:ProblemWorkflow}
\end{figure}

Figure \ref{fig:ProblemWorkflow} describes the problem workflow. For each time step $t$, a weather map with six bands (or channels) acts as the input image. More details on the data set are provided in Section \ref{sec:Data}. The model analyzes input images  to estimate solar and wind power production. In the first CNN or NN models, one input at time $t$ leads to one solar and one wind power production value for time $t$. To include temporal effects, we also investigate the models that take five stacked sequential weather maps as input, leading to an input image with 30 channels. Finally, in recurrent neural network techniques such as LSTM, the model receives both an input weather map at time $t$ and information from previous weather maps before time $t$ through so-called hidden data \cite{Hochreiter97LSTMoriginal}.  

As mentioned, the long-term goal is to forecast renewable power production, e.g. 1 day or 1 week ahead. In that set-up, the model would forecast energy production at time $t+\Delta t_f$ using weather data at time $t$, where $\Delta t_f$ is the forecast horizon. Implicitly, the model would need to learn to estimate power outputs and predict the effect of future weather patterns. The latter is a complex task by itself. Fortunately, it is also a well established field. Hence, we divide the production forecasting task into two distinct problems, each with a specialized model: (i) a weather forecasting model and (ii) a power estimation model. This work focuses on the latter. Future work can then leverage the presented models to forecast solar and wind power production using forecasted weather maps as inputs into the power estimation model. Separating the weather forecast problem from the power estimation problem also allows for a more flexible framework. As the power estimation model does not need to understand local weather patterns, it requires less adjustments to be applied to diverse locations. For example, this can be achieved using transfer learning where only a small subset of the models layers are adjusted.

\section{California Weather Data Set} \label{sec:Data}
A secondary yet significant contribution is the dataset. To train the model, we require input weather maps and output renewable energy production from the same time period.

\begin{figure}[htbp]
\begin{center}
   \includegraphics[width=.4\linewidth]{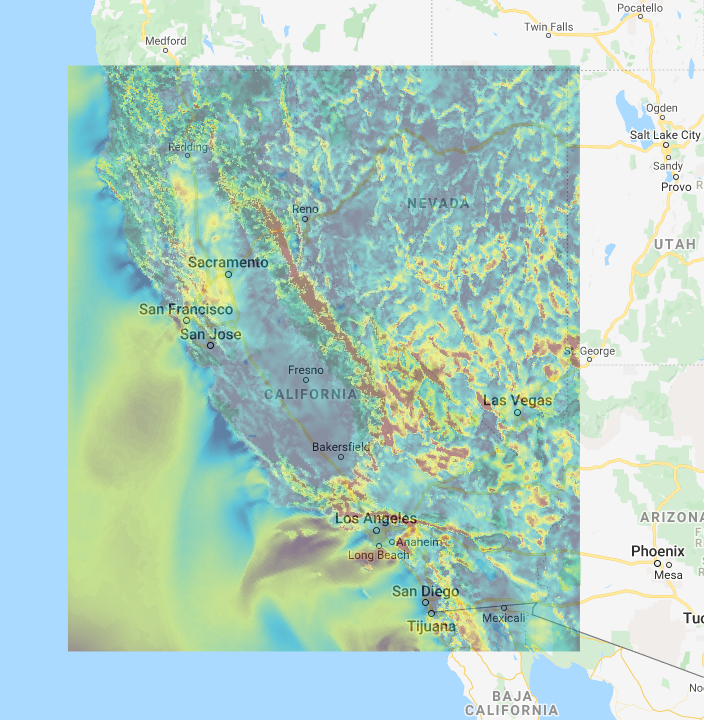}
\end{center}
   \caption{A snapshot of the data overlaying a topographic map.}
\label{fig:MapData}
\end{figure}

The weather data is extracted from a National Weather Service (NWS) and NOAA real-time mesoscale analysis \cite{WeatherData}. The data has a temporal resolution of 1 hour and spatial resolution of $2.5$km$\times 2.5$km. Via the Google Earth Engine, we crop the data to a bounding box around the state of California (Figure \ref{fig:MapData}) and between 01-01-2019 to 31-12-2019. When extracting, we set the coordinate reference system to \textit{NAD 83 / California Albers} (Code EPSG:3310) to ensure the pixels are exported as a straight-standing rectangle similar to the one selected on the map. Due to the earth’s curvature, we lose a couple of pixels in the corners that will be treated in pre-processing.

\begin{figure}[htbp]
\centering
\begin{subfigure}{.3\linewidth}
   \centering
   \includegraphics[width=\linewidth]{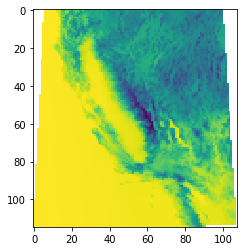}
   \caption{}
\end{subfigure}
\begin{subfigure}{.3\linewidth}
   \centering
   \includegraphics[width=\linewidth]{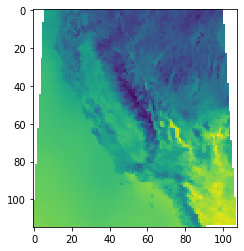}
   \caption{}
\end{subfigure}
\begin{subfigure}{.3\linewidth}
   \centering
   \includegraphics[width=\linewidth]{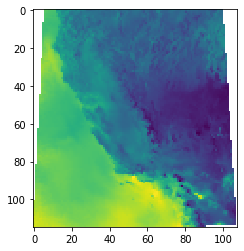}
   \caption{}
\end{subfigure}
\\
\begin{subfigure}{.3\linewidth}
   \centering
   \includegraphics[width=\linewidth]{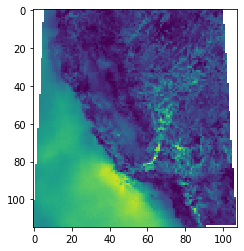}
   \caption{}
\end{subfigure}
\begin{subfigure}{.3\linewidth}
   \centering
   \includegraphics[width=\linewidth]{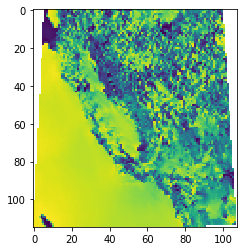}
   \caption{}
\end{subfigure}
\begin{subfigure}{.3\linewidth}
   \centering
   \includegraphics[width=\linewidth]{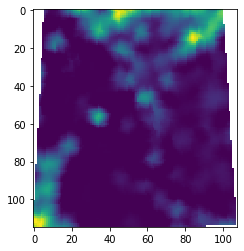}
   \caption{}
\end{subfigure}
 \caption{The z-score normalized weather input channels of a single training sample. The parameters displayed are (a) pressure, (b) temperature, (c) humidity, (d) wind speed, (e) wind direction, and (f) cloud cover. Note that each pixel of each weather parameter only has a single value and that the colors displayed are only for visualization purposes (yellow to blue for high to low). The axis values indicate pixel count in each direction.}
 \label{fig:WeatherImages}
\end{figure}

To be able to efficiently use the images, the data is down-scaled and normalized as follows. First, the six (out of 13) most important bands are selected: pressure (Pa), temperature ($^{\circ}$C), humidity(kg/kg), wind speed(m/s), wind direction($^\circ$ where North is 0$^\circ$), and cloud cover(\%). To manage the size of the data set, we coarsen the pixels to a spatial dimension of $10$km$\times10$km. Next, the bands are z-score normalized over the whole data set such that they each have mean $\mu = 0$ and standard deviation $\sigma = 1$. The resulting input images are $6\times115\times108$ [channels$\times$pixel height$\times$pixel width] and are displayed for one time sample in Figure \ref{fig:WeatherImages}. Finally, we set the corner pixels equal to 0. 

The renewable production data is sourced  from the California ISO website and is specified by source and reported every five minutes. To match the temporal scale of the corresponding weather images, we average the energy production output over an hour. We only focus on solar and wind energy as they are unpredictable and very sensitive to weather. Other renewable energy sources, such as geothermal or hydro, are generally controllable and are predominantly governed by capacity and demand. 

The dataset and code routines are publicly available. Moreover, we encourage researchers to further improve the proposed model as well as investigate different regions by generating new weather data images from the NWS data in GEE.

\subsection{Training and Testing}
The dataset is randomly shuffled, and divided into a training, validation, and test set following a 80\%-10\%-10\% split of samples, respectively. Table \ref {table:DataSetSize} specifies the number of samples in each data subset. Considering the size of the dataset, indices of the samples in each subset are stored and efficient data loaders are designed to extract samples, allowing the use of the same train-val-test indices for both the single input and stacked input set-ups. The only exception is that we exclude the first five samples (first five hours on Jan 1, 2019) when using stacked inputs, because these stacked samples are incomplete. 

\begin{table}[ht]
\begin{center}
\caption{Sample count in training, validation, and test sets.}
\label{table:DataSetSize}
\begin{tabular}{ c c c }
 \hline 
  Training Set & Validation Set & Test Set \\ [0.5ex] 
 \hline
 7008 & 876 & 875 \\ 
 \hline
\end{tabular}
\end{center}
\end{table}

In the current set up, we do not preserve chronological order of training and testing data. However, as temporal features (e.g. day, month, hour) are not included in the inputs, the model can not identify subsequent steps. Therefore the violation of chronology is acceptable and allows us to test on all 4 seasons instead of only testing in the last season, i.e. winter in our dataset. Future work will look into retraining and retesting as time progresses, using chronologically ordered data. We note that temporal features are often very valuable for time series forecasting. As such, their omission potentially makes the current forecasting problem significantly harder. We expect to see improved results when including them in future work.


\section{Models}
In this paper, we present a reference linear model and a ResNet inspired convolutional model. DenseNets and LSTM-CNN models are also briefly discussed, though not displayed, because their current performance was mediocre. To compare the models fairly, we keep many design choices the same for both the models. Their performance is evaluated both with traditional loss values and specifically designed accuracy metrics. The implementation is done using the publicly available Pytorch library \cite{pytorch}.

\subsection{Loss Function and Optimization}
The final layer in each model outputs a power value for solar and wind energy. As these output signals are continuous variables, losses for the estimations are computed using root mean squared errors ($RMSE$):
\begin{equation}
RMSE = \sqrt{\frac{1}{n}\sum_{i=1}^{n}(y_{i} - \hat{y_{i}})^{2}}. \label{RMSE}
\end{equation}
where $n$ is number of samples, $y_{i}$ is the $i$th data point, and $\hat{y_{i}}$ is the $i$th data point estimation.

To minimize the loss, the model weights are updated and optimized using backpropagation and the ADAM algorithm \cite{kingma2014adam}. ADAM is chosen because it exhibited the most consistent loss evolution. We  recognize that ADAM can converge to a sharp local minima that may be far away from the global minimum. Optimizing with more-sensitive yet less robust optimizers, such as Stochastic Gradient Descent + Momentum, is the subject of future work. Here, we aim to present respectable results, showing the potential of our proposed methodology.

All models are trained for a total of 20 epochs in four stages. We define an epoch as a pass through the entire training data set, i.e. $7008$ samples. Every 5 epochs, a new stage is initiated, where the learning rate is decreased to allow finer adjustments of the model. A new stage is initiated when the RMSE loss stops decreasing. L2 regularization is applied to reduce overfitting. After hyperparameter tuning we choose regularization weights of $\lambda = 0.01$ and $\lambda = 0.001$ for the linear and ResNet models, respectively. During training, we use batch sizes of 16 samples. 

\subsection{Accuracy Metric}
In addition to the RMSE loss, we design a more meaningful accuracy metric to qualitatively evaluate the performance. To this end, we assess an absolute error relative to the average power output, $\Bar{y}_{\text{train}}$, in the training set, i.e.

\begin{equation}
    \text{Acc}_i = 1- \frac{|\hat{y}_i - y_i|}{\Bar{y}_{\text{train}}}, \label{eq:accuracy}
\end{equation}

where $\hat{y}_i$ is the predicted power output and $y_i$ is the actual power output for sample $i$. Note this metric is computed separately for solar and wind energy because the mean power outputs of the training data are $3.02$ MW and $1.68$ MW, respectively. 

\subsection{Linear Neural Network Model} \label{sec:linearNeuralNetworkModel}
We design a linear model that starts wide and gradually narrows to the two estimation nodes in four layers--800, 400, 200, and 2 nodes. The input image is flattened and then connected to the first fully connected block. A block consists of a fully connected layer preceded by a dropout layer with a dropout probability equal to 0.2 and followed by a ReLU activation function. Note that we also add an unconventional final ReLU activation layer after the last fully connected layer. This addition ensures that outputs are strictly positive and helps the model capture the zero solar output at night. We experienced significant improvements in prediction accuracy by applying this final ReLU activation layer. Note that this strategy does deactivate a large share of the network at initialization. Hence we require dropout layers to enable nodes to be activated during training. 

\subsection{ResNet-inspired Convolutional Model}
\begin{figure}[htbp]
\begin{center}
  \includegraphics[width=.7\linewidth]{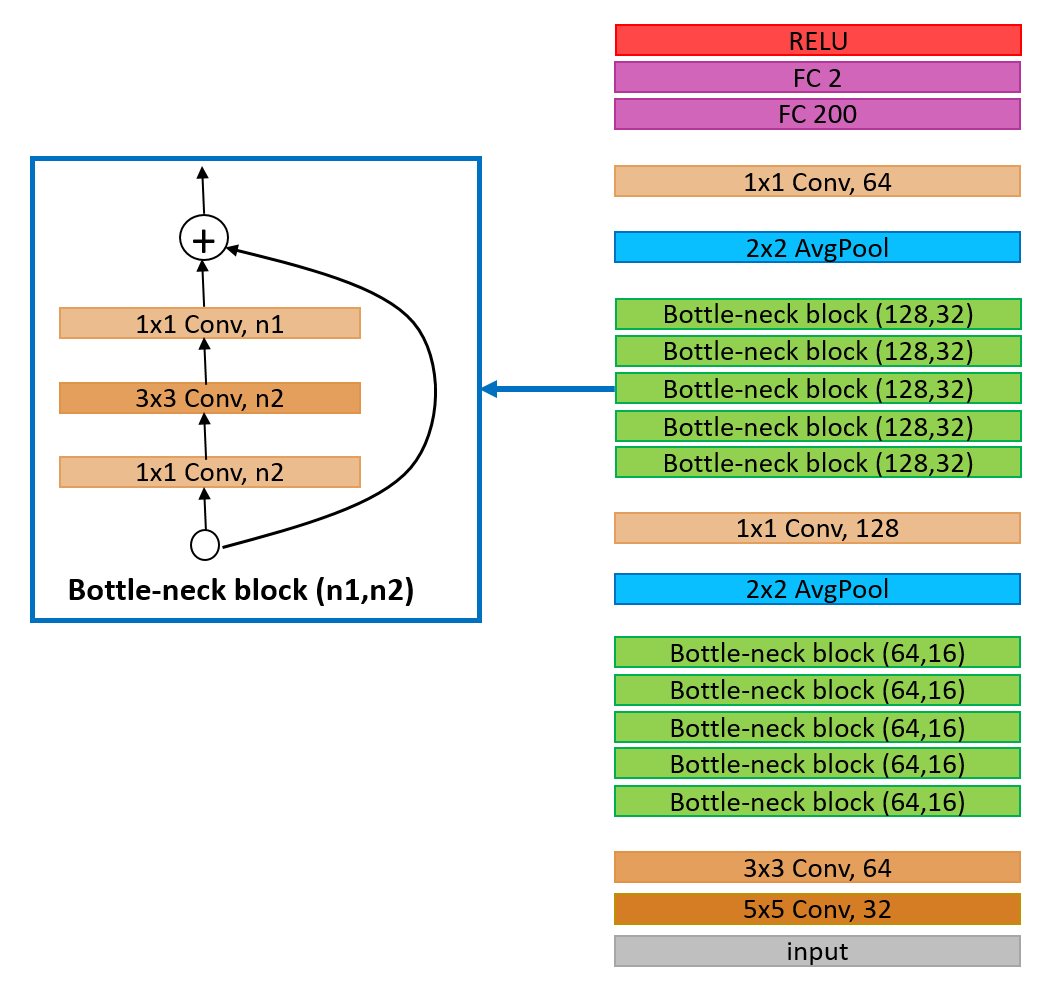}
\end{center}
  \caption{ResNet-inspired Model.}
\label{fig:ResNetModel}
\end{figure}

A ResNet inspired model is designed to take advantage of the spatial correlation in the data. The designed model consists of 38 layers, including 10 "bottle-neck" building blocks \cite{He_2016_CVPR_ResNet}. These building blocks and the overall network are presented in Figure \ref{fig:ResNetModel}. Every convolutional layer is followed by a batch normalization and a ReLU activation function. Additionally, average pooling layers are included to reduce the image dimensions. Finally, the model's last layers are two fully connected layers and a final ReLU layer. These follow the same reasoning as presented in Section \ref{sec:linearNeuralNetworkModel}. Furthermore, we precise that all convolutional layers are initialized using Kaiming initialization \cite{KaimingInitialization}, linear layers are initialized with a uniform random distribution, and bias terms are set to 0. We point out that although the ResNet-inspired convolutional model has significantly more layers, it does not have more parameters than the presented linear model.

\subsection{Other Models}
A DenseNet model and a CNN-LSTM model were also investigated. DenseNet did not prove more effective than the ResNet-inspired model and therefore was not included  in this report. The initial results of the CNN-LSTM model were mediocre, partly due to data and computational power shortage. Future work may attempt to unlock higher accuracy with these time-series specialized methods using greater resources.



\section{Results}
The models' performances on the 2019 California data set are summarized and compared in the following subsections. We evaluate the loss and accuracy to understand the prediction results. Additionally, saliency maps are presented to understand and validate the models' estimations.

The overall accuracy is presented in Table \ref{table:SolarAcc} and Table \ref{table:WindAcc}. The ResNet model with five stacked sequential weather maps as input outperforms all the other models in both wind and solar power estimation. The other three models exchange the rest of the ranks depending on energy source and data subset. The following subsections explain the reasons for these performances.

\begin{table}[ht]
\begin{center}
\caption{Model Solar Power Accuracy.}
\label{table:SolarAcc}
\begin{tabular}{ l c c c }
 \hline 
     & Training & Validation & Test \\  
 \hline
    Single Linear & 0.9319 & 0.8673 & 0.8645 \\ 
    Single ResNet & 0.9335 & 0.8350 & 0.8340 \\ 
    Stacked Linear & 0.9208 & 0.8736 & 0.8813 \\ 
    Stacked ResNet & \textbf{0.9414} & \textbf{0.8882} & \textbf{0.8840} \\ 
 \hline
\end{tabular}
\end{center}
\end{table}

\begin{table}[ht]
\begin{center}
\caption{Model Wind Power Accuracy.}
\label{table:WindAcc}
\begin{tabular}{ l c c c }
 \hline 
     & Training & Validation & Test \\ 
 \hline
    Single Linear & 0.8800 & 0.8469 & 0.8454 \\ 
    Single ResNet & 0.9025 & 0.8426 & 0.8433 \\ 
    Stacked Linear & 0.8360 & 0.8103 & 0.8189 \\ 
    Stacked ResNet & \textbf{0.9037} & \textbf{0.8668} & \textbf{0.8713} \\ 
 \hline
\end{tabular}
\end{center}
\end{table}

\subsection{Loss Evolution}
\begin{figure}[htbp]
\begin{center}
  \includegraphics[width=.6\linewidth]{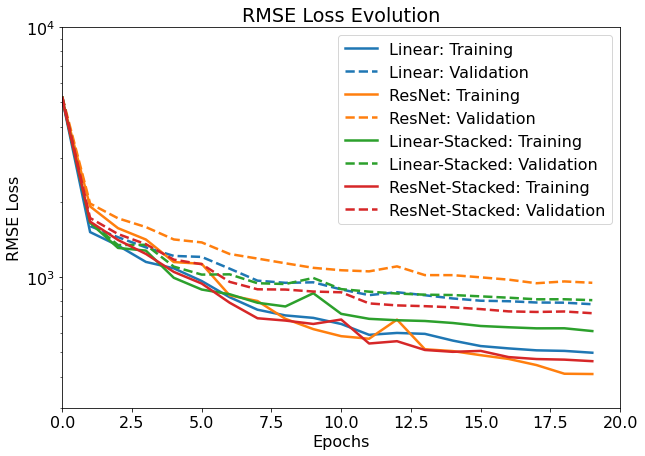}
\end{center}
  \caption{Loss evolution for all models during training. Losses are evaluated on the training and validation data subsets separately.}
\label{fig:Loss}
\end{figure}

Figure \ref{fig:Loss} displays the RMSE loss evolution on the full training and validation data sets while the model is being trained. Both ResNet models perform very well on the training data. However, the single input model overfits and has poor validation loss. High regularization weights or increased dropout fails to  improve its performance. Similarly, the validation-training gap for the single input linear model is larger than the gap with a stacked input. In this case the validation losses are comparable. The following subsections explain these discrepancies by investigating the solar and wind power predictions. 

\subsection{Solar Power Estimation Accuracy}
\begin{figure}[htbp]
\begin{center}
  \includegraphics[width=.6\linewidth]{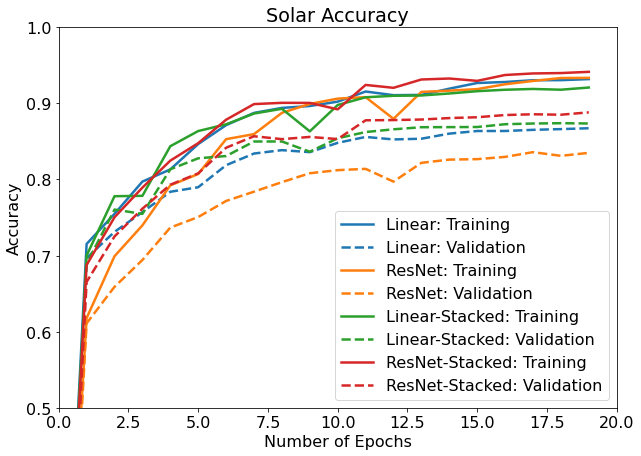}
\end{center}
  \caption{Solar power estimation accuracy (see Eq. \eqref{eq:accuracy}) evolution while training for each of the different models.}
\label{fig:SolarAccuracy}
\end{figure}

\begin{figure}[htbp]
\centering
\begin{subfigure}{.8\linewidth}
   \centering
   \includegraphics[width=\linewidth]{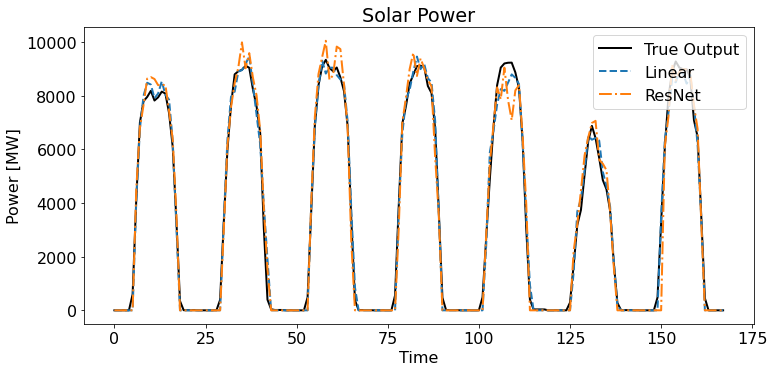}
   \caption{Single input models}
\end{subfigure}
\\
\begin{subfigure}{.8\linewidth}
   \centering
  \includegraphics[width=\linewidth]{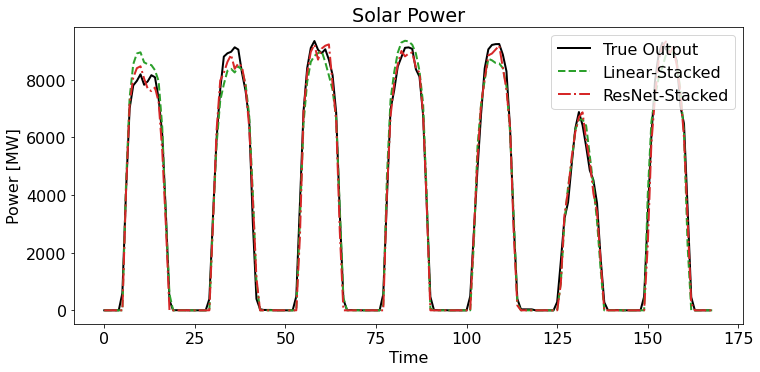}
   \caption{Stacked input models}
\end{subfigure}
  \caption{Solar energy comparison between true power output and model estimations. The data reflects the week of May 5 - May 11, 2019. Note the data is displayed chronologically for visualization purposes. Training and testing is performed with a random sampling of data points. As such, the visualization includes training, validation, and testing data samples.}
\label{fig:SolarPrediction}
\end{figure}

We analyze the models' solar power estimates by plotting the evolution of our accuracy metric, see eq.\eqref{eq:accuracy}, in Figure \ref{fig:SolarAccuracy} and the power estimates for a chosen week as shown in Figure \ref{fig:SolarPrediction}. 

The solar accuracy evolution shows similar behavior to the RMSE loss. The most notable difference is that the stacked input models exhibit better accuracy for the same loss value. Especially interesting is how well the linear models perform. Since solar intensity is a highly local quantity, the model’s main objective is locating important solar locations (with the exception of forecasting incoming cloud cover). Because spatially correlated information is not required, the linear model is adequately suited. 
Comparing the true solar output to the various model estimates in Figure \ref{fig:SolarPrediction}, we see that the linear models indeed perform close to on par with the convolutional models. On the other hand, the plots do show a distinct difference between single input models and stacked input models. In particular, the single input models exaggerate the high frequency variations at high power output while the stacked input models retain a smoother production profile.

\subsection{Wind Power Estimation Accuracy}
\begin{figure}[htbp]
\begin{center}
  \includegraphics[width=.6\linewidth]{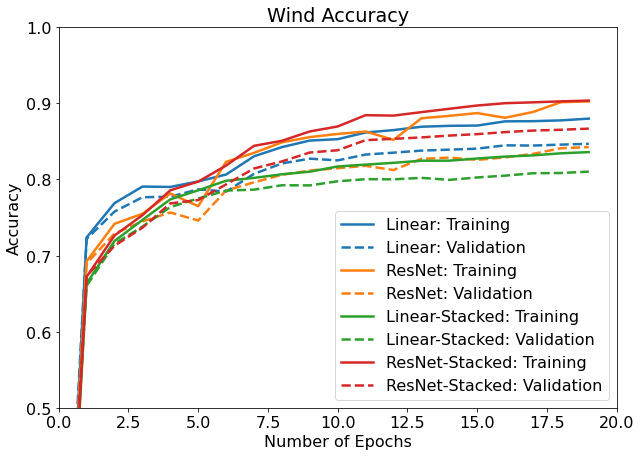}
\end{center}
  \caption{Wind power estimation accuracy (see Eq. \eqref{eq:accuracy}) evolution while training for each of the different models.}
\label{fig:WindAccuracy}
\end{figure}

Wind power is a more complex energy source and less temporally predictable, making the wind power estimation problem more challenging. As such, the models obtain lower accuracy scores for wind power than obtained for solar power. Table \ref{table:WindAcc} summarizes the wind power estimation results. Moreover, we find that convolutional models are more effective than linear models at estimating wind power production. To support this finding we plot the accuracy curves in Figure \ref{fig:WindAccuracy}. This result makes sense considering the importance of spatially correlated data when estimating wind power. Overall, the stacked ResNet model clearly outperforms the other models in validation and testing accuracy. 

\begin{figure}[htbp]
\centering
\begin{subfigure}{.8\linewidth}
   \centering
  \includegraphics[width=\linewidth]{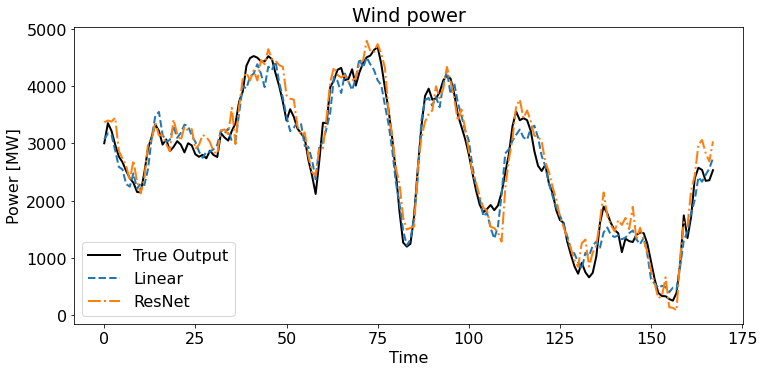}
   \caption{Single input models}
\end{subfigure}
\\
\begin{subfigure}{.8\linewidth}
   \centering
  \includegraphics[width=\linewidth]{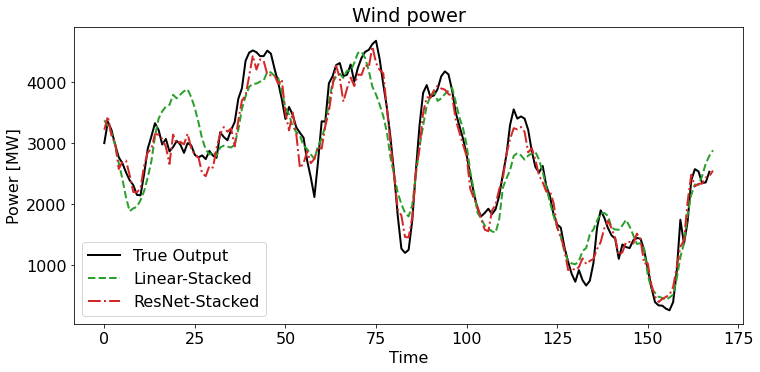}
   \caption{Stacked input models}
\end{subfigure}
  \caption{Wind energy comparisons between true power output and model estimations.The data reflects the week of May 5 - May 11, 2019. Note the data is displayed chronologically for visualization purposes. Training and testing is performed with a random sampling of data points. As such, the visualization includes training, validation, and testing data samples.}
\label{fig:WindPrediction}
\end{figure}

Figure \ref{fig:WindPrediction} displays the wind power estimates for the first full week of May in 2019. Again, we see that the stacked input models present much smoother predictions. However, in this case, we see that the stacked linear model exaggerates the smoothing. As a consequence, this model loses high frequency details and struggles to capture the unpredictable magnitude of low-frequency temporal oscillations. On the other hand, the single input models display very peaky estimation curves which, although accurate, are not very realistic. The stacked ResNet model combines the best of both worlds and outputs an accurate and realistic representation of the true wind power output.  

\subsection{Saliency Maps}
Saliency maps visualize which parts of an input image influence the model's prediction. To obtain a saliency map, we compute the absolute value of the model output’s gradient with respect to the model input and then take the maximum value over the channels to represent the pixel intensity \cite{simonyan2013deep}.

\begin{figure}[htbp]
\centering
\begin{subfigure}{.4\linewidth}
   \centering
   \includegraphics[width=\linewidth]{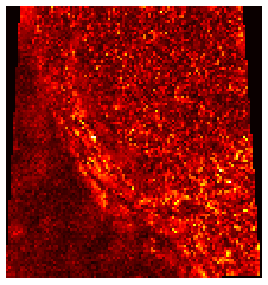}
   \caption{Solar 12pm: (noon)}
\end{subfigure}
\begin{subfigure}{.4\linewidth}
   \centering
   \includegraphics[width=\linewidth]{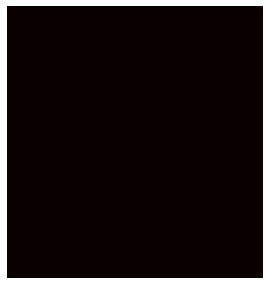}
   \caption{Solar 22:00}
\end{subfigure}
\begin{subfigure}{.4\linewidth}
   \centering
   \includegraphics[width=\linewidth]{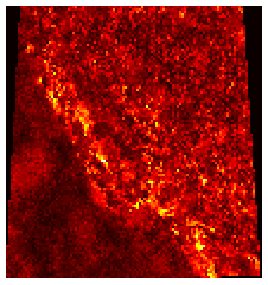}
   \caption{Wind 12:00 (noon)}
\end{subfigure}
\begin{subfigure}{.4\linewidth}
   \centering
   \includegraphics[width=\linewidth]{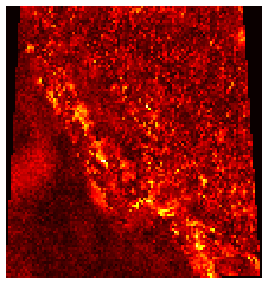}
   \caption{Wind 22:00}
 \end{subfigure}
 \caption{Saliency maps for two different samples: May 8 at 12:00 (noon) and 22:00.}
 \label{fig:SaliencyMaps}
\end{figure}

\begin{figure}[htbp]
\begin{center}
  \includegraphics[width=.5 \linewidth]{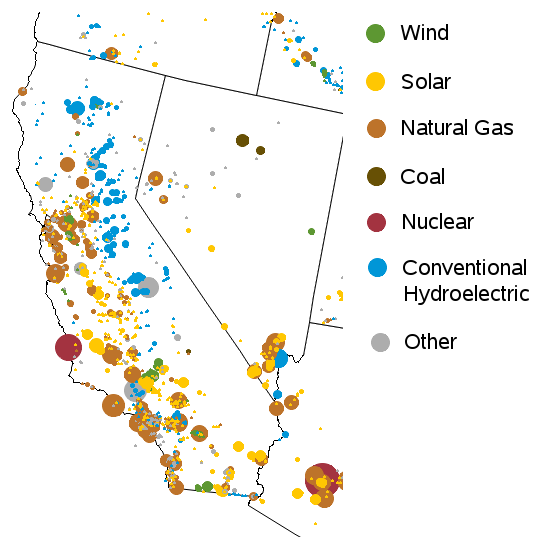}
\end{center}
  \caption{Locations of Electricity Producers in California in May 2020 colored by source. The size of a dot is proportional to the capacity of the plant. \cite{EIA}}
\label{fig:ElecProdLoc}
\end{figure}

Figure \ref{fig:SaliencyMaps} presents saliency maps of two samples at different times of day. As expected, we see that the solar saliency map at 22:00 is black because there is no sun and the output is equal to 0. On the other hand, we see a semicircular pattern of high intensity at noon, starting near the north central valley, passing by LA, and bending towards Las Vegas. Comparing this pattern with the location of solar plants displayed in Figure \ref{fig:ElecProdLoc}, we see a clear correlation. Moreover, both wind and solar saliency maps correctly show low intensity in the ocean areas that are of little importance. Note that this probably would have been different if we would have had one model perform both the forecasting and power estimation task (see Section \ref{sec:ProblemDescription}). In that scenario, weather patterns over the ocean can have an important influence on future weather patterns on land.

The wind saliency maps also show bright spots at corresponding locations to the wind power plants. These bright spots are larger and also cover populated areas. The increased spatial dimension corresponds to the importance of a larger spatial area around a wind power plant. The explanation as to why the intersection with locations of high population density is less straightforward. One possibility is that wind energy is subjected to curtailment depending on demand. Hence, the model might use information about weather conditions of surrounding cities to implicitly estimate energy consumption, which in turn would influence the energy production estimate.

\subsection{Discussion \& Future Work}
As this is an initial exploration of this application, the results can be further improved by adjusting model parameters, overcoming data limitations, and seeking more complex models.

The saliency maps display a reasonable amount of low amplitude noise. We expect a near-perfect model to display the same targeted bright spots without any noise in unimportant regions. Moreover, the maps also hint at the importance of other parameters, such as demand. To maximize model accuracy, more input data streams should be added. Temporal features, demand forecasts and sun incidence angle are examples of good candidates. A more diverse set of inputs will allow the model to more effectively extract the share of useful information that is encoded in weather maps. 

\begin{figure}[htbp]
\centering
\begin{subfigure}{.4\linewidth}
   \centering
   \includegraphics[width=\linewidth]{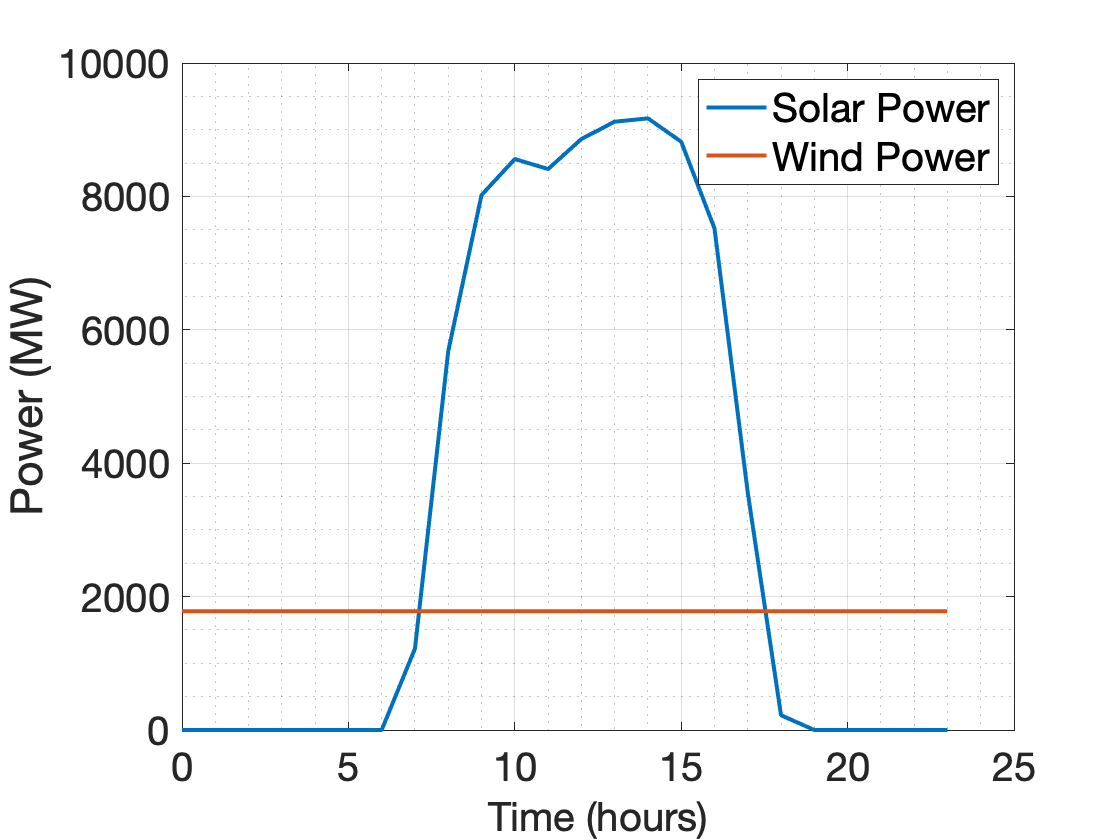}
   \caption{October $6^{th}$ 2019}
\end{subfigure}
\begin{subfigure}{.4\linewidth}
   \centering
   \includegraphics[width=\linewidth]{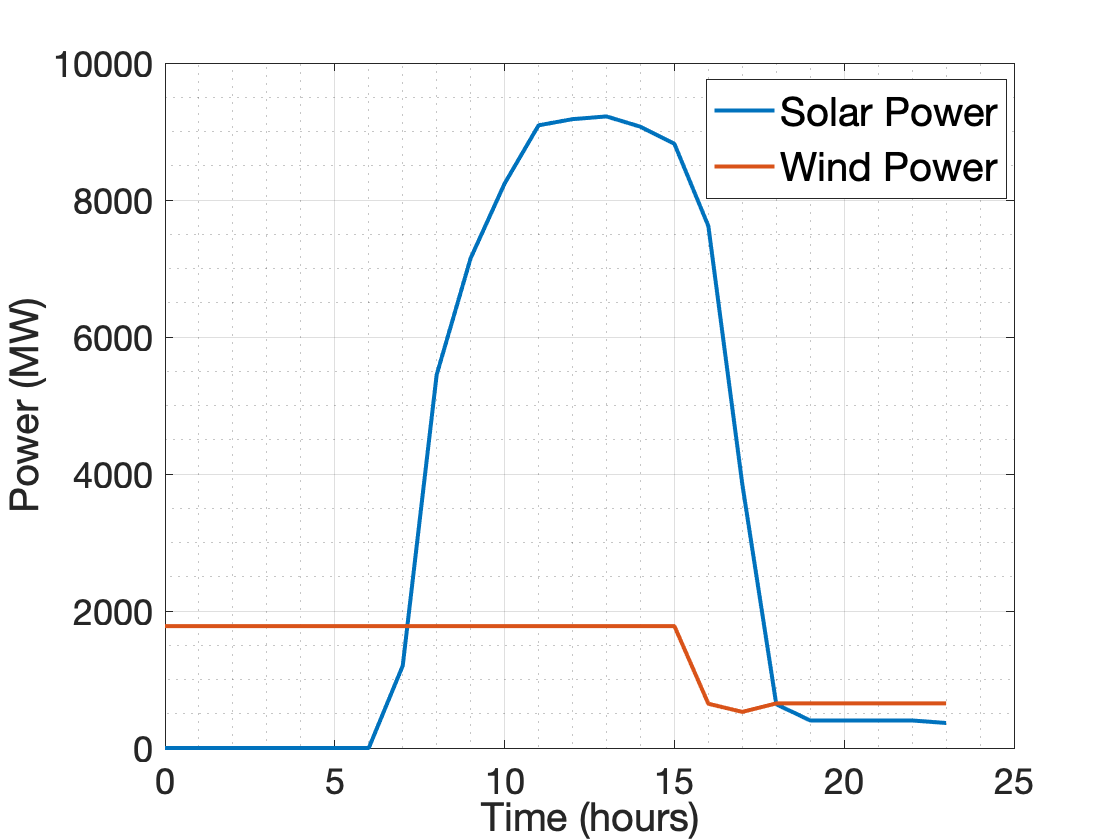}
   \caption{October $7^{th}$ 2019}
\end{subfigure}
\caption{Solar and wind power for days with incorrectly reported wind energy.}
\label{fig:oddData}
\end{figure}

The production dataset used in this work contains two days where the wind power production data is reported as constant. The authors presume this is false or through human intervention (e.g. curtailment). Figure \ref{fig:oddData} exhibits the solar and wind power graphs for October $6^{th}$ and $7^{th}$, 2019. These false data points negatively impact the models' accuracies. To resolve this, one can opt to exclude these data points. We decided to include them because the incorrect data points are few in number and only relate to wind power -- the solar power data for those times was still valuable. Additionally, by including them, the dataset remains complete and chronologically intact.  A better strategy might be to use a model to generate synthetic data for October 6 and 7. As such, one would avoid obtaining false correlations and preserve all valid data.

\section{Conclusion \& Future Work}
This work takes a first step towards a novel approach in forecasting renewable energy production and, in the bigger picture, any large-scale energy market variable. To this end, we propose using state-of-the-art computer vision techniques on weather maps to estimate renewable energy production. Our ResNet-inspired model outperforms other traditional deep learning techniques and obtains accuracies close to 90\% for both solar and wind power. We also show the value of using multiple input frames from times leading up to the estimation. The underlying reasons for the model’s superior prediction capabilities are highlighted with an analysis of the output signal characteristics for different types of energy, and with saliency maps that indicate the precise locations that influence the model’s prediction. Here, we were able to verify that the model independently assigns importance to locations with renewable energy production facilities. 

This work is not an end, but rather a beginning of a novel strategy to forecast energy market variables. Future work can apply the power estimation model to weather forecasts to obtain wind and solar power forecasts. We also emphasize that the results presented in this work are achieved using solely weather data as input. Additional non-image data, such as temporal data or sun incidence angle, should be included as inputs to further improve performance and more accurately leverage weather map data.

The presented methodology can also be applied to other important society scale variables. For example, energy demand is closely correlated to weather because of indoor climate control. In this line of thinking, one could envision a set up similar to state-of-the-art image classification model infrastructure. That is, a framework where sub-models predict lower level variables which are then used as input in the global model to predict the primary variable. For example, a model that identifies humans in a picture might have sub-models identifying faces, arms and legs. In particular, forecasts of complex market variables, such as energy price, might benefit significantly from a multi-level structure.

Another interesting tangential avenue could be to explore how changes in climate can impact renewable energy generation. To do so, one could estimate renewable energy production on altered weather data (e.g. increased temperature, higher humidity or less cloud cover).

Finally, future work naturally includes enhancing the presented models. This work serves as a proof-of-concept and we encourage others to extend our research. In this, alternative architectures and models are of interest, e.g. CNN-LSTM or temporal convolutional network (TCN). Additionally, higher resolution input data should be explored and new locations of interest can be evaluated. Transfer learning and conducting training/testing at different locations would also be valuable topics of future work in going towards real-world deployment.

\section*{Acknowledgements}
The authors thank Jon Braatz and Stace Maples for their support in collecting the weather data. We also thank Robyn Lockwood for her great help in revising the manuscript. Finally, we thank Titiaan Palazzi, Patrick Sheehan and Pieter Verhoeven for the interesting discussions that led to this work.
Sebastian Bosma is supported by a named Stanford Graduate Fellowship in Science and Engineering (SGF).

\section*{Contributions}
The model and methods were designed by SB and NN. Hyperparameter tuning was performed by SB and NN. Weather data was collected by SB. Power data was collected by NN. SB and NN wrote the paper.

{
\bibliography{CNNbib}
}

\end{document}